\documentclass[11pt,a4paper,english,nofootinbib]{revtex4}
\usepackage{lmodern}

\usepackage[T1]{fontenc}
\usepackage[latin9]{inputenc}
\usepackage{babel}
\usepackage{amsmath}
\usepackage{amssymb}
\usepackage{esint}
\usepackage[unicode=true,pdfusetitle,
 bookmarks=true,bookmarksnumbered=false,bookmarksopen=false,
 breaklinks=false,pdfborder={0 0 1},backref=false,colorlinks=false]
 {hyperref}

\makeatletter

\pdfpageheight\paperheight
\pdfpagewidth\paperwidth

\@ifundefined{textcolor}{}
{%
 \definecolor{BLACK}{gray}{0}
 \definecolor{WHITE}{gray}{1}
 \definecolor{RED}{rgb}{1,0,0}
 \definecolor{GREEN}{rgb}{0,1,0}
 \definecolor{BLUE}{rgb}{0,0,1}
 \definecolor{CYAN}{cmyk}{1,0,0,0}
 \definecolor{MAGENTA}{cmyk}{0,1,0,0}
 \definecolor{YELLOW}{cmyk}{0,0,1,0}
 }

\usepackage{latexsym}\usepackage{bm}

\makeatother

\begin{document}

\title{Spectra, vacua and the unitarity of Lovelock gravity in $D$-dimensional
AdS spacetimes}

\author{Tahsin Ça\u{g}r\i{} \c{S}i\c{s}man}

\email{sisman@metu.edu.tr}

\selectlanguage{english}%

\affiliation{Department of Physics,\\
 Middle East Technical University, 06531, Ankara, Turkey}

\author{\.{I}brahim Güllü }

\email{e075555@metu.edu.tr}

\selectlanguage{english}%

\affiliation{Department of Physics,\\
 Middle East Technical University, 06531, Ankara, Turkey}

\author{Bayram Tekin}

\email{btekin@metu.edu.tr}

\selectlanguage{english}%

\affiliation{Department of Physics,\\
 Middle East Technical University, 06531, Ankara, Turkey}

\date{\today}
\begin{abstract}
We explicitly confirm the expectation that generic Lovelock gravity
in $D$ dimensions has a unitary massless spin-2 excitation around
any one of its constant curvature vacua just like the cosmological
Einstein gravity. The propagator of the theory reduces to that of
Einstein's gravity, but scattering amplitudes must be computed with
an effective Newton's constant which we provide. Tree-level unitarity
imposes a single constraint on the parameters of the theory yielding
a wide range of unitary region. As an example, we explicitly work
out the details of the cubic Lovelock theory. 
\end{abstract}
\maketitle

\section{Introduction}

In $D$ spacetime dimensions, Lovelock gravity \cite{Lovelock1,Lovelock2}
is defined by the Lagrangian density 
\begin{equation}
\mathcal{L}_{\text{Lovelock}}\left(R_{\rho\sigma}^{\mu\nu}\right)=\sum_{n=0}^{\left[\frac{D}{2}\right]}a_{n}\mathcal{L}_{n},\label{eq:Lovelock}
\end{equation}
 where $a_{n}$'s are dimensionful constants, $\left[\frac{D}{2}\right]$
corresponds to the integer part of $\frac{D}{2}$, and all the indices
on the tensors run from $\left(0,\dots,D-1\right)$. Each term in
the full Lagrangian density is given as 
\begin{equation}
\mathcal{L}_{n}=\delta_{\nu_{1}\dots\nu_{2n}}^{\mu_{1}\dots\mu_{2n}}\prod_{p=1}^{n}R_{\mu_{2p-1}\mu_{2p}}^{\nu_{2p-1}\nu_{2p}},\label{eq:Lovelock_order}
\end{equation}
 where $\delta_{\nu_{1}\dots\nu_{2n}}^{\mu_{1}\dots\mu_{2n}}$ is
the generalized Kronecker delta or the determinant tensor defined
as usual in terms of the Kronecker deltas; 
\begin{equation}
\delta_{\nu_{1}\dots\nu_{2n}}^{\mu_{1}\dots\mu_{2n}}\equiv\det\left|\begin{array}{ccc}
\delta_{\nu_{1}}^{\mu_{1}} & \dots & \delta_{\nu_{1}}^{\mu_{2n}}\\
\vdots & \ddots & \vdots\\
\delta_{\nu_{2n}}^{\mu_{1}} & \dots & \delta_{\nu_{2n}}^{\mu_{2n}}
\end{array}\right|.\label{eq:gen_kronecker}
\end{equation}
 This theory is presumably the most natural generalization of Einstein's
gravity in $D$ dimensions with the well-known property that the field
equations are second order in the derivatives of the metric tensor
(in fact, this is one of the defining properties of the theory). The
lowest order term with $n=0$ corresponds to the cosmological constant
which is followed by the Einstein-Hilbert action with $n=1$ and the
Gauss-Bonnet (GB) combination with $n=2$. GB combination appears
in low energy string theory dictated by supersymmetry \cite{Zwiebach,Boulware_Deser}.
As a result of general covariance of the action, the field equations
are also covariantly divergence free. Moreover, Lovelock gravity is
the only higher derivative theory that does not suffer the Buchdahl's
inequality \cite{Buchdahl} that exists between the metric formulation
and the Palatini formulation--which assumes a generic connection \emph{a
priori}--of higher derivative gravity theories \cite{Exirifard}.
This result is quite interesting, since it yields a dynamical derivation
of equivalence principle for a class of torsion-free theories \cite{Exirifard,Sheikh-Jabbari}.
Another property of (\ref{eq:Lovelock}) is that for \emph{even} $D$,
the highest-order term does not contribute to the field equations,
since its variation is a total derivative (i.e. $\mathcal{L}_{D/2}$
is a topological invariant in even $D$ dimensions). There is a nontrivial
point here: In the first order formalism with vielbeins and the spin
connection, one can explicitly show that $\mathcal{L}_{D/2}$ can
be written as a boundary term. But, in the metric formulation there
is no natural covariant vector made of the metric tensor $g_{\mu\nu}$
and its derivatives $\partial_{\rho}g_{\mu\nu}$; therefore, to see
that the $\frac{D}{2}$ term does not contribute to the field equations,
one proceeds by showing that under arbitrary variations of the metric,
that term yields a total divergence. For example, see \cite{tHooft}
for the case of the GB combination. If one does not insist that $\mathcal{L}_{D/2}$
be written as a boundary term in a covariant way, one can find noncovariant
expressions. For example, see \cite{Padmanabhan} where an algorithm
of constructing boundary terms is given and as an example the Einstein-Hilbert
Lagrangian density in two dimensions is explicitly written as a boundary
term in a \emph{noncovariant} way.

Recently \cite{Franklin}, canonical analysis of Lovelock theory,
more specifically $D=5$ Einstein-GB theory, was carried out via the
ADM decomposition \cite{ADM}. Several other properties such as cosmological
solutions \cite{Deruelle}, black hole solutions \cite{BTZ}, thermodynamical
properties \cite{Padmanabhan-Thermo1,Padmanabhan-Thermo2,Kastor}
of Lovelock theories have been studied.

In this work, we expand the Lovelock action around one of its (anti)-de
Sitter {[}(A)dS{]} vacua up to $O\left(h^{2}\right)$ in the metric
perturbation where $h_{\mu\nu}\equiv g_{\mu\nu}-\bar{g}_{\mu\nu}$
to study the propagator structure, perturbative spectrum and the unitarity.
As expected, we find that Lovelock gravity has only a massless spin-2
excitation in its spectrum just like the (cosmological) Einstein theory.
The free propagator of Lovelock theory reduces to that of cosmological
Einstein's gravity with an \emph{effective} Newton's constant and
an \emph{effective} cosmological constant. The main result of this
work is to provide how these two effective constants can be computed
in terms of the constants, $a_{n}$, in the Lovelock action, and discuss
the conditions on the parameters coming from the tree-level unitarity
requirement.

The layout of this paper is follows: in Section II, we introduce the
Lovelock gravity, find its $\left[\frac{D}{2}\right]$ constant curvature
vacua, find an equivalent action whose propagator matches the propagator
of Lovelock gravity, and discuss the tree-level unitarity of the theory.
As the first nontrivial example, we work out the spectrum and the
vacua of the cubic Lovelock gravity. Most of the details of the computations
are delegated to the Appendix.

\section{Propagator Structure of the Lovelock Action}

To study the perturbative spectrum of Lovelock gravity (\ref{eq:Lovelock})
around one of its constant curvature vacua, naively one should find
the field equations and linearize them around their maximally symmetric
solutions. But, this route is somewhat complicated because of the
complexity of the action. Instead, let us use the \emph{equivalent
quadratic action technique}, which was employed in several works before
\cite{Hindawi,Gullu-BIUnitarity,Gullu-Cubic3D,Sisman-AllUni}. Since
the technique is described in more detail in these works, here let
us briefly recapitulate how it works. Suppose one would like to find
the constant curvature vacua and excitations around the vacua, namely
the propagator of a generic gravity action that is constructed from
the metric and the contractions of the Riemann tensor. The form of
the Lagrangian density action can be taken as $\mathcal{L}\equiv\sqrt{-g}F\left(R_{\rho\sigma}^{\mu\nu}\right)$.
The question is to find an equivalent quadratic Lagrangian density
$\mathcal{L}_{\text{quad-equal}}\equiv\sqrt{-g}f_{\text{quad-equal}}\left(R_{\rho\sigma}^{\mu\nu}\right)$
whose $O\left(h\right)$, representing the maximally symmetric vacua,
and $O\left(h^{2}\right)$, representing the propagator, expansions
match that of the original Lagrangian. The equivalent quadratic action
can be found from the curvature expansion around yet to be found maximally
symmetric vacua with the Riemann tensor $\bar{R}_{\rho\sigma}^{\mu\nu}=\frac{2\Lambda}{\left(D-1\right)\left(D-2\right)}\left(\delta_{\rho}^{\mu}\delta_{\sigma}^{\nu}-\delta_{\sigma}^{\mu}\delta_{\rho}^{\nu}\right)$
as 
\begin{equation}
f_{\text{quad-equal}}\left(R_{\rho\sigma}^{\mu\nu}\right)\equiv\sum_{i=0}^{2}\left[\frac{\partial^{i}F}{\partial\left(R_{\rho\sigma}^{\mu\nu}\right)^{i}}\right]_{\bar{R}_{\rho\sigma}^{\mu\nu}}\left(R_{\rho\sigma}^{\mu\nu}-\bar{R}_{\rho\sigma}^{\mu\nu}\right)^{i}.\label{eq:f_quad}
\end{equation}
 The allowed values for $\Lambda$ comes from the order $O\left(h\right)$
expansion of the original action, but more directly they can be obtained
from the maximally symmetric vacuum equation of the equivalent quadratic
theory.

It is not difficult to see%
\footnote{The $i=2$ term in the curvature expansion involves totally antisymmetric
contractions of two Riemann tensors giving the GB combination.%
} that the equivalent quadratic action for Lovelock gravity should
be in the Einstein-GB form 
\begin{equation}
\mathcal{L}_{{\rm EGB}}=\frac{1}{\kappa}\left(R-2\Lambda_{0}\right)+\gamma\,\chi_{\text{GB}},\qquad\chi_{\text{GB}}\equiv R_{\mu\nu\rho\sigma}R^{\mu\nu\rho\sigma}-4R_{\mu\nu}R^{\mu\nu}+R^{2}\label{eq:EGB}
\end{equation}
 What is of course remarkable is that the free theory, which is determined
by $O\left(h\right)$ and $O\left(h^{2}\right)$ expansion, of (\ref{eq:EGB})
also matches that of cosmological Einstein gravity $\mathcal{L}_{E}=\frac{1}{\kappa_{e}}\left(R-2\Lambda\right)$
with the following effective parameters \cite{Sisman-AllUni} 
\begin{equation}
\frac{1}{\kappa_{e}}=\frac{1}{\kappa}+\frac{4\Lambda\left(D-3\right)\left(D-4\right)}{\left(D-1\right)\left(D-2\right)}\gamma,\qquad\frac{\Lambda-\Lambda_{0}}{2\kappa}+\gamma\frac{\left(D-3\right)\left(D-4\right)}{\left(D-1\right)\left(D-2\right)}\Lambda^{2}=0.\label{eq:Equiv_Eins_params}
\end{equation}

The main problem is then to relate $\kappa_{e}$ and $\Lambda$ to
the parameters $a_{n}$ in the Lovelock action. In order to find these
relations, we need to compute (\ref{eq:f_quad}) or more explicitly
the following quantities 
\begin{equation}
\mathcal{L}_{\text{Lovelock}}\left(\bar{R}_{\rho\sigma}^{\mu\nu}\right),\qquad\left[\frac{\partial\mathcal{L}_{\text{Lovelock}}}{\partial R_{\rho\sigma}^{\mu\nu}}\right]_{0}\left(R_{\rho\sigma}^{\mu\nu}-\bar{R}_{\rho\sigma}^{\mu\nu}\right),\qquad\left[\frac{\partial^{2}\mathcal{L}_{\text{Lovelock}}}{\partial R_{\rho\sigma}^{\mu\nu}\partial R_{\alpha\beta}^{\lambda\gamma}}\right]_{0}\left(R_{\rho\sigma}^{\mu\nu}-\bar{R}_{\rho\sigma}^{\mu\nu}\right)\left(R_{\alpha\beta}^{\lambda\gamma}-\bar{R}_{\alpha\beta}^{\lambda\gamma}\right).\label{eq:Lovelock_equiv_quant}
\end{equation}
 From now on the subindex {}``$0$'' means that the corresponding
quantity is evaluated in the background. Details of these calculations
are given in the Appendix, here we simply write the final expressions:
\begin{equation}
\mathcal{L}_{\text{Lovelock}}\left(\bar{R}_{\rho\sigma}^{\mu\nu}\right)=D!\sum_{n=0}^{\left[\frac{D}{2}\right]}a_{n}\left[\frac{4\Lambda}{\left(D-1\right)\left(D-2\right)}\right]^{n}\frac{1}{\left(D-2n\right)!},
\end{equation}
 
\begin{equation}
\left[\frac{\partial\mathcal{L}_{\text{Lovelock}}}{\partial R_{\rho\sigma}^{\mu\nu}}\right]_{0}\left(R_{\rho\sigma}^{\mu\nu}-\bar{R}_{\rho\sigma}^{\mu\nu}\right)=\sum_{n=0}^{\left[\frac{D}{2}\right]}a_{n}n\left[\frac{4\Lambda}{\left(D-1\right)\left(D-2\right)}\right]^{n-1}\frac{2\left(D-2\right)!}{\left(D-2n\right)!}\left(R-\frac{2D\Lambda}{D-2}\right),
\end{equation}
 
\begin{multline}
\left[\frac{\partial^{2}\mathcal{L}_{\text{Lovelock}}}{\partial R_{\rho\sigma}^{\mu\nu}\partial R_{\alpha\beta}^{\lambda\gamma}}\right]_{0}\left(R_{\rho\sigma}^{\mu\nu}-\bar{R}_{\rho\sigma}^{\mu\nu}\right)\left(R_{\alpha\beta}^{\lambda\gamma}-\bar{R}_{\alpha\beta}^{\lambda\gamma}\right)\\
=4\sum_{n=0}^{\left[\frac{D}{2}\right]}a_{n}n\left(n-1\right)\left[\frac{4\Lambda}{\left(D-1\right)\left(D-2\right)}\right]^{n-2}\frac{\left(D-4\right)!}{\left(D-2n\right)!}\chi_{\text{GB}}\\
-2\sum_{n=0}^{\left[\frac{D}{2}\right]}a_{n}n\left(n-1\right)\left[\frac{4\Lambda}{\left(D-1\right)\left(D-2\right)}\right]^{n-1}\frac{2\left(D-2\right)!}{\left(D-2n\right)!}\left(R-\frac{D\Lambda}{D-2}\right).
\end{multline}
 As a result, the equivalent quadratic action that has the same $O\left(h\right)$
and $O\left(h^{2}\right)$ expansions%
\footnote{In fact, $O\left(h^{0}\right)$ of the equivalent quadratic Lagrangian
also gives the same order of the Lovelock theory, but this is an irrelevant
constant.%
} with the Lovelock theory (\ref{eq:Lovelock}) can be constructed
from the equivalent quadratic Lagrangian density 
\begin{equation}
f_{\text{quad-equal}}\left(R_{\rho\sigma}^{\mu\nu}\right)=-2\left(D-2\right)!\sum_{n=0}^{\left[\frac{D}{2}\right]}\tilde{a}_{n}\left[R-\frac{\left(n-1\right)D\Lambda}{n\left(D-2\right)}-\frac{\left(n-1\right)}{4\Lambda\left(n-2\right)}\frac{\left(D-1\right)}{\left(D-3\right)}\chi_{\text{GB}}\right],\label{eq:Lovelock_equiv_quad_Lag}
\end{equation}
 where $\tilde{a}_{n}$ is defined as 
\[
\tilde{a}_{n}\equiv a_{n}\frac{n\left(n-2\right)}{\left(D-2n\right)!}\left[\frac{4\Lambda}{\left(D-1\right)\left(D-2\right)}\right]^{n-1}.
\]
 Here, $\tilde{a}_{n}$ vanishes for $n=2$, but because of the $\left(n-2\right)$
term in the denominator the contribution does not vanish. The propagator
of (\ref{eq:Lovelock}) matches that of (\ref{eq:Lovelock_equiv_quad_Lag})
which itself has exactly the same propagator as the cosmological Einstein's
theory but the Newton's constant is modified. Therefore, 
\begin{equation}
\mathcal{L}_{{\rm Lovelock}}\left(h^{2}\right)=-\frac{1}{2\kappa_{e}}h^{\mu\nu}\mathcal{D}_{\mu\nu\alpha\beta}^{E}h^{\alpha\beta},
\end{equation}
 where $\mathcal{D}_{\mu\nu\alpha\beta}^{E}$ is the propagator of
the cosmological Einstein theory which propagates a unitary massless
spin-2 particle as long as $\kappa_{e}>0$ \cite{Porrati,GulluTekin}.

Note that $n=0,1,2$ terms of (\ref{eq:Lovelock_equiv_quad_Lag})
give the cosmological constant, the Ricci scalar and the GB combination
as expected. The first nontrivial term comes from the cubic Lovelock
term whose explicit form is 
\begin{align}
\frac{\mathcal{L}_{3}}{8}= & -8R^{\mu\nu\rho\sigma}R_{\mu\phantom{\tau}\rho}^{\phantom{\mu}\tau\phantom{\rho}\gamma}R_{\nu\tau\sigma\gamma}+4R^{\mu\nu\rho\sigma}R_{\mu\nu}^{\phantom{\mu\nu}\tau\gamma}R_{\rho\sigma\tau\gamma}-24R^{\mu\nu}R_{\phantom{\rho\sigma\tau}\mu}^{\rho\sigma\tau}R_{\rho\sigma\tau\nu}\nonumber \\
 & +3RR^{\mu\nu\rho\sigma}R_{\mu\nu\rho\sigma}+24R^{\mu\nu}R^{\rho\sigma}R_{\mu\rho\nu\sigma}+16R^{\mu\nu}R_{\mu}^{\rho}R_{\nu\rho}-12RR^{\mu\nu}R_{\mu\nu}+R^{3}.\label{eq:ED6}
\end{align}
 Since no homogeneous in curvature Lagrangian density can have a nonzero
maximally symmetric vacuum, together with $\mathcal{L}_{3}$ one needs
to consider at least one of the lower order Lovelock terms. Here,
for the sake of generality, we consider all the lower order terms
together with $\mathcal{L}_{3}$. Then, the equivalent quadratic action
to $\mathcal{L}=a_{0}+a_{1}\mathcal{L}_{1}+a_{2}\mathcal{L}_{2}+a_{3}\mathcal{L}_{3}$
follows as 
\begin{align}
f_{\text{quad-equal}}\left(R_{\rho\sigma}^{\mu\nu}\right)= & a_{0}+2a_{1}R+4a_{2}\chi_{\text{GB}}\nonumber \\
 & -96a_{3}\frac{\left(D-3\right)\left(D-4\right)\left(D-5\right)\Lambda^{2}}{\left(D-1\right)^{2}\left(D-2\right)}\left(R-\frac{2D\Lambda}{3\left(D-2\right)}\right)\label{eq:EQA-Lovelock_to_cubic}\\
 & +48a_{3}\frac{\left(D-4\right)\left(D-5\right)\Lambda}{\left(D-1\right)\left(D-2\right)}\chi_{\text{GB}}.\nonumber 
\end{align}
 As expected, for $D=3,4\text{ and }5$, the terms coming from $\mathcal{L}_{3}$
explicitly vanish. For $D=6$, they do not vanish explicitly, but
as we show below they do not contribute to the field equations which
is consistent with the fact that $\mathcal{L}_{3}$ is a topological
invariant in $D=6$. This is also true for $\mathcal{L}_{2}$: for
$D=4$ $a_{2}$ does not contribute to the field equations. {[}Note
that for $D=3$, $\mathcal{L}_{2}$ also does not contribute to the
field equations because the GB combination is identically zero.{]}
Similarly, $a_{1}$ does not contribute to the field equations in
$D=2$.

Let us re-write the equivalent quadratic Lagrangian (\ref{eq:Lovelock_equiv_quad_Lag})
of the Lovelock theory as 
\begin{equation}
f_{\text{quad-equal}}=\frac{1}{\tilde{\kappa}}\left(R-2\tilde{\Lambda}_{0}\right)+\tilde{\gamma}\,\chi_{\text{GB}},
\end{equation}
 where the parameters are defined as 
\begin{equation}
\frac{1}{\tilde{\kappa}}\equiv-2\left(D-2\right)!\sum_{n=0}^{\left[\frac{D}{2}\right]}a_{n}\frac{n\left(n-2\right)}{\left(D-2n\right)!}\left[\frac{4\Lambda}{\left(D-1\right)\left(D-2\right)}\right]^{n-1},\label{eq:k_tilde}
\end{equation}
 
\begin{equation}
\frac{\tilde{\Lambda}_{0}}{\tilde{\kappa}}\equiv-\frac{D!}{4}\sum_{n=0}^{\left[\frac{D}{2}\right]}a_{n}\frac{\left(n-1\right)\left(n-2\right)}{\left(D-2n\right)!}\left[\frac{4\Lambda}{\left(D-1\right)\left(D-2\right)}\right]^{n},\label{eq:L_tilde}
\end{equation}
 
\begin{equation}
\tilde{\gamma}\equiv2\left(D-4\right)!\sum_{n=0}^{\left[\frac{D}{2}\right]}a_{n}\frac{n\left(n-1\right)}{\left(D-2n\right)!}\left[\frac{4\Lambda}{\left(D-1\right)\left(D-2\right)}\right]^{n-2}.\label{eq:g_tilde}
\end{equation}
 One can have a further reduction in representing the maximally symmetric
vacua and the free part of the Lovelock theory with an equivalent
action by using the fact discussed above: the vacua and the propagator
of the Einstein-GB theory can be represented with cosmological Einstein's
gravity having the modified parameters given in (\ref{eq:Equiv_Eins_params}).
Then, we can write 
\begin{equation}
I_{{\rm equal-Lovelock}}=\int d^{D}x\sqrt{-g}\frac{1}{\kappa_{e}}\left(R-2\Lambda\right),
\end{equation}
 where by using (\ref{eq:Equiv_Eins_params}) and (\ref{eq:k_tilde}-\ref{eq:g_tilde}),
one can find that$\Lambda$ satisfies the vacuum equation 
\begin{equation}
0=\sum_{n=0}^{\left[\frac{D}{2}\right]}a_{n}\frac{\left(D-2n\right)}{\left(D-2n\right)!}\left[\frac{4}{\left(D-1\right)\left(D-2\right)}\right]^{n}\Lambda^{n},\label{eq:Vacuum}
\end{equation}
 and $\kappa_{e}$ becomes 
\begin{align}
\frac{1}{\kappa_{e}}= & 2\left(D-3\right)!\sum_{n=0}^{\left[\frac{D}{2}\right]}a_{n}\frac{n\left(D-2n\right)}{\left(D-2n\right)!}\left[\frac{4\Lambda}{\left(D-1\right)\left(D-2\right)}\right]^{n-1}.\label{eq:k_eff-Lovelock}
\end{align}
 As we discussed, since $\mathcal{L}_{D/2}$ is a topological invariant
for \emph{even }$D$, its contribution to the field equations should
vanish for that dimension; and this fact can be verified by observing
the appearance of $\left(D-2n\right)$ factor in (\ref{eq:Vacuum})
and (\ref{eq:k_eff-Lovelock}). Positivity of $\kappa_{e}$ is the
single constraint to have unitary massless spin-2 excitation which
simply gives a bound on one of the $a_{n}$'s in terms of the others.
Reality of $\Lambda$, in general, also puts a constraint on the parameters.
Similar results have been found in the first order formalism of Lovelock
gravity in two different ways: first, by expanding the action up to
second order in one of the components of $h_{\mu\nu}$ and second
by using the spherically symmetric black hole solution in AdS backgrounds
\cite{Edelstein,Camanho}.

\section{Conclusion}

We have found the $O\left(h^{2}\right)$ expansion of generic Lovelock
gravity in $D$ dimensions around one of its constant curvature vacua,
and explicitly confirmed the expectation that just like Einstein's
theory, there is a single massless spin-2 excitation%
\footnote{Besides Lovelock gravity there are other higher curvature theories
which propagate just a single massless spin-2 excitation around (A)dS
backgrounds, see for example the critical gravity \cite{LuPope,DeserLiu}. %
}. Qualitatively, Lovelock theory is built to have this property, but
actual computation of the effective Newton's constant and the effective
cosmological constant was lacking which was remedied above. We have
given the cubic curvature Lovelock action as an explicit nontrivial
example. Our construction is based on the fact that the propagator
and the maximally symmetric vacua for any gravity theory whose action
involves the contractions of the Riemann tensor can be obtained from
an equivalent quadratic action.

Even though we have concentrated on the perturbative spectrum and
the vacua, it is not difficult to see that from our construction conserved
gravitational charges such as energy and angular momenta of black
holes in asymptotically (anti)-de Sitter spacetimes can be found for
Lovelock gravity by following the procedure of \cite{DeserTekinPRL,DeserTekin}:
the expression for the charge will be just like the one in the cosmological
Einstein-Hilbert theory with the effective Newton's constant and the
effective cosmological constant.

\section{Acknowledgments}

This work is supported by the T{Ü}B\.{I}TAK Grant No. 110T339. We
would like to thank José D. Edelstein for useful discussions.

\appendix

\section{Terms of the Equivalent Action}

In the calculations involving the generalized Kronecker delta $\delta_{\nu_{1}\dots\nu_{2n}}^{\mu_{1}\dots\mu_{2n}}$,
which is nothing but a determinantal form, we frequently use the following
relation for an $n\times n$ matrix $A$ 
\begin{equation}
\det A=\epsilon_{\alpha_{1}\dots\alpha_{n}}A_{\alpha_{1}1}A_{\alpha_{2}2}\dots A_{\alpha_{n}n},\label{eq:detA}
\end{equation}
 where the convention for the permutation symbol is $\epsilon_{12\dots2n}=+1$.
At the risk of being pedantic, let us explicitly obtain the form of
$\delta_{\nu_{1}\dots\nu_{2n}}^{\mu_{1}\dots\mu_{2n}}$ with the help
of (\ref{eq:detA}). First, consider the elements of the following
$2n\times2n$ matrix 
\begin{equation}
L=\left(\begin{array}{ccc}
\delta_{\nu_{1}}^{\mu_{1}} & \dots & \delta_{\nu_{1}}^{\mu_{2n}}\\
\vdots & \ddots & \vdots\\
\delta_{\nu_{2n}}^{\mu_{1}} & \dots & \delta_{\nu_{2n}}^{\mu_{2n}}
\end{array}\right),
\end{equation}
 where the index $\nu$ counts the rows, and the index $\mu$ counts
the columns; i.e. one has $L_{ij}=\delta_{\nu_{i}}^{\mu_{j}}$, and
for example, $L_{\alpha_{1}1}=\delta_{\nu_{\alpha_{1}}}^{\mu_{1}}$.
Then, one can write $\delta_{\nu_{1}\dots\nu_{2n}}^{\mu_{1}\dots\mu_{2n}}$
as 
\begin{equation}
\delta_{\nu_{1}\dots\nu_{2n}}^{\mu_{1}\dots\mu_{2n}}=\epsilon_{\alpha_{1}\dots\alpha_{2n}}\delta_{\nu_{\alpha_{1}}}^{\mu_{1}}\delta_{\nu_{\alpha_{2}}}^{\mu_{2}}\dots\delta_{\nu_{\alpha_{2n}}}^{\mu_{2n}}.\label{eq:Gen_Kro-Del}
\end{equation}
 Here, note that $2n$ should be smaller than the dimension of the
spacetime $D$, but need not to be equal to $D$.

Now, let us discuss how the term $\delta_{\nu_{1}\dots\nu_{2k}\nu_{2k+1}\dots\nu_{2n}}^{\mu_{1}\dots\mu_{2k}\nu_{2k+1}\dots\nu_{2n}}$
is related to $\delta_{\nu_{1}\dots\nu_{2k}}^{\mu_{1}\dots\mu_{2k}}$.
Using (\ref{eq:Gen_Kro-Del}), one can find how $n\rightarrow n$
case is related to the $n\rightarrow n-\frac{1}{2}$ case 
\begin{equation}
\delta_{\nu_{1}\dots\nu_{2k}\nu_{2k+1}\dots\nu_{2n}}^{\mu_{1}\dots\mu_{2k}\nu_{2k+1}\dots\nu_{2n}}=\left[D-\left(2n-1\right)\right]\epsilon_{\alpha_{1}\alpha_{2}\dots\alpha_{2n-1}}\delta_{\nu_{\alpha_{1}}}^{\mu_{1}}\delta_{\nu_{\alpha_{2}}}^{\mu_{2}}\dots\delta_{\nu_{\alpha_{2k}}}^{\mu_{2k}}\delta_{\nu_{\alpha_{2k+1}}}^{\nu_{2k+1}}\dots\delta_{\nu_{\alpha_{2n-1}}}^{\nu_{2n-1}}.
\end{equation}
 Using this recursive relation, it is possible to obtain the desired
result which will be sufficient in the computation of the equivalent
quadratic action 
\begin{equation}
\delta_{\nu_{1}\dots\nu_{2k}\nu_{2k+1}\dots\nu_{2n}}^{\mu_{1}\dots\mu_{2k}\nu_{2k+1}\dots\nu_{2n}}=\frac{\left(D-2k\right)!}{\left(D-2n\right)!}\delta_{\nu_{1}\dots\nu_{2k}}^{\mu_{1}\dots\mu_{2k}}.\label{eq:Recursion}
\end{equation}

\subsection{Zeroth Order}

Let us calculate $\mathcal{L}_{\text{Lovelock}}\left(\bar{R}_{\rho\sigma}^{\mu\nu}\right)$
which has the form 
\begin{equation}
\mathcal{L}_{\text{Lovelock}}\left(\bar{R}_{\rho\sigma}^{\mu\nu}\right)=\sum_{n=0}^{\left[\frac{D}{2}\right]}a_{n}\delta_{\nu_{1}\dots\nu_{2n}}^{\mu_{1}\dots\mu_{2n}}\prod_{p=1}^{n}\bar{R}_{\mu_{2p-1}\mu_{2p}}^{\nu_{2p-1}\nu_{2p}}.
\end{equation}
 By using 
\begin{equation}
\bar{R}_{\rho\sigma}^{\mu\nu}=\frac{2\Lambda}{\left(D-1\right)\left(D-2\right)}\left(\delta_{\rho}^{\mu}\delta_{\sigma}^{\nu}-\delta_{\sigma}^{\mu}\delta_{\rho}^{\nu}\right),\label{eq:AdS}
\end{equation}
 one has 
\begin{equation}
\delta_{\nu_{1}\dots\nu_{2n}}^{\mu_{1}\dots\mu_{2n}}\prod_{p=1}^{n}\bar{R}_{\mu_{2p-1}\mu_{2p}}^{\nu_{2p-1}\nu_{2p}}=\left[\frac{4\Lambda}{\left(D-1\right)\left(D-2\right)}\right]^{n}\delta_{\nu_{1}\dots\nu_{2n}}^{\nu_{1}\dots\nu_{2n}}=\left[\frac{4\Lambda}{\left(D-1\right)\left(D-2\right)}\right]^{n}\frac{D!}{\left(D-2n\right)!},\label{eq:Zeroth_order_contractions}
\end{equation}
 where the second equality follows from (\ref{eq:Recursion}). Note
that the value of this form for $n_{max}=\left[\frac{D}{2}\right]$
is same for both even and odd dimensions. Then, $\mathcal{L}_{\text{Lovelock}}\left(\bar{R}_{\rho\sigma}^{\mu\nu}\right)$
becomes 
\begin{equation}
\mathcal{L}_{\text{Lovelock}}\left(\bar{R}_{\rho\sigma}^{\mu\nu}\right)=D!\sum_{n=0}^{\left[\frac{D}{2}\right]}a_{n}\left[\frac{4\Lambda}{\left(D-1\right)\left(D-2\right)}\right]^{n}\frac{1}{\left(D-2n\right)!}.
\end{equation}

\subsection{First order}

The first order term in the equivalent quadratic action has the form
\begin{align}
\left[\frac{\partial\mathcal{L}_{\text{Lovelock}}}{\partial R_{\rho\sigma}^{\mu\nu}}\right]_{0}\left(R_{\rho\sigma}^{\mu\nu}-\bar{R}_{\rho\sigma}^{\mu\nu}\right)= & \sum_{n=0}^{\left[\frac{D}{2}\right]}a_{n}\delta_{\nu_{1}\dots\nu_{2n}}^{\mu_{1}\dots\mu_{2n}}\sum_{r=1}^{n}\left(\prod_{\underset{\left(p\ne r\right)}{p=1}}^{n}\bar{R}_{\mu_{2p-1}\mu_{2p}}^{\nu_{2p-1}\nu_{2p}}\right)R_{\mu_{2r-1}\mu_{2r}}^{\nu_{2r-1}\nu_{2r}}\nonumber \\
 & -\sum_{n=0}^{\left[\frac{D}{2}\right]}a_{n}\delta_{\nu_{1}\dots\nu_{2n}}^{\mu_{1}\dots\mu_{2n}}n\left(\prod_{p=1}^{n}\bar{R}_{\mu_{2p-1}\mu_{2p}}^{\nu_{2p-1}\nu_{2p}}\right),
\end{align}
 where the term in the second line is calculated in (\ref{eq:Zeroth_order_contractions});
and after use of (\ref{eq:AdS}), the term in the first line becomes
\begin{align}
\delta_{\nu_{1}\dots\nu_{2n}}^{\mu_{1}\dots\mu_{2n}}\sum_{r=1}^{n}\left(\prod_{\underset{\left(p\ne r\right)}{p=1}}^{n}\bar{R}_{\mu_{2p-1}\mu_{2p}}^{\nu_{2p-1}\nu_{2p}}\right)R_{\mu_{2r-1}\mu_{2r}}^{\nu_{2r-1}\nu_{2r}}= & \left[\frac{4\Lambda}{\left(D-1\right)\left(D-2\right)}\right]^{n-1}n\delta_{\nu_{1}\nu_{2}\nu_{3}\dots\nu_{2n}}^{\mu_{1}\mu_{2}\nu_{3}\dots\nu_{2n}}R_{\mu_{1}\mu_{2}}^{\nu_{1}\nu_{2}}.
\end{align}
 Using (\ref{eq:Recursion}), one can further reduce this form to
\begin{equation}
\delta_{\nu_{1}\dots\nu_{2n}}^{\mu_{1}\dots\mu_{2n}}\sum_{r=1}^{n}\left(\prod_{\underset{\left(p\ne r\right)}{p=1}}^{n}\bar{R}_{\mu_{2p-1}\mu_{2p}}^{\nu_{2p-1}\nu_{2p}}\right)R_{\mu_{2r-1}\mu_{2r}}^{\nu_{2r-1}\nu_{2r}}=\left[\frac{4\Lambda}{\left(D-1\right)\left(D-2\right)}\right]^{n-1}2n\frac{\left(D-2\right)!}{\left(D-2n\right)!}R.\label{eq:First_order_contractions}
\end{equation}
 This result together with (\ref{eq:Zeroth_order_contractions}) yields
the first order term of the equivalent quadratic action as 
\begin{equation}
\left[\frac{\partial\mathcal{L}_{\text{Lovelock}}}{\partial R_{\rho\sigma}^{\mu\nu}}\right]_{0}\left(R_{\rho\sigma}^{\mu\nu}-\bar{R}_{\rho\sigma}^{\mu\nu}\right)=\sum_{n=0}^{\left[\frac{D}{2}\right]}a_{n}n\left[\frac{4\Lambda}{\left(D-1\right)\left(D-2\right)}\right]^{n-1}\frac{2\left(D-2\right)!}{\left(D-2n\right)!}\left(R-\frac{2D\Lambda}{D-2}\right).
\end{equation}

\subsection{Second order}

The second order term in the equivalent quadratic action has the form
\begin{multline}
\left[\frac{\partial^{2}\mathcal{L}_{\text{Lovelock}}}{\partial R_{\rho\sigma}^{\mu\nu}\partial R_{\alpha\beta}^{\lambda\gamma}}\right]_{0}\left(R_{\rho\sigma}^{\mu\nu}-\bar{R}_{\rho\sigma}^{\mu\nu}\right)\left(R_{\alpha\beta}^{\lambda\gamma}-\bar{R}_{\alpha\beta}^{\lambda\gamma}\right)\\
=\sum_{n=0}^{\left[\frac{D}{2}\right]}a_{n}\delta_{\nu_{1}\dots\nu_{2n}}^{\mu_{1}\dots\mu_{2n}}\sum_{r=1}^{n}\sum_{\underset{\left(r\ne q\right)}{q=1}}^{n}\left(\prod_{\underset{\left(p\ne r,q\right)}{p=1}}^{n}\bar{R}_{\mu_{2p-1}\mu_{2p}}^{\nu_{2p-1}\nu_{2p}}\right)R_{\mu_{2r-1}\mu_{2r}}^{\nu_{2r-1}\nu_{2r}}R_{\mu_{2q-1}\mu_{2q}}^{\nu_{2q-1}\nu_{2q}}\\
-\sum_{n=0}^{\left[\frac{D}{2}\right]}a_{n}\delta_{\nu_{1}\dots\nu_{2n}}^{\mu_{1}\dots\mu_{2n}}2\left(n-1\right)\sum_{r=1}^{n}\left(\prod_{\underset{\left(p\ne r\right)}{p=1}}^{n}\bar{R}_{\mu_{2p-1}\mu_{2p}}^{\nu_{2p-1}\nu_{2p}}\right)R_{\mu_{2r-1}\mu_{2r}}^{\nu_{2r-1}\nu_{2r}}\\
\sum_{n=0}^{\left[\frac{D}{2}\right]}a_{n}\delta_{\nu_{1}\dots\nu_{2n}}^{\mu_{1}\dots\mu_{2n}}n\left(n-1\right)\left(\prod_{p=1}^{n}\bar{R}_{\mu_{2p-1}\mu_{2p}}^{\nu_{2p-1}\nu_{2p}}\right),
\end{multline}
 where the second and the third terms on the right-hand side are calculated
in (\ref{eq:First_order_contractions}) and (\ref{eq:Zeroth_order_contractions}),
respectively. On the other hand, the first term takes the form 
\begin{multline}
\delta_{\nu_{1}\dots\nu_{2n}}^{\mu_{1}\dots\mu_{2n}}\sum_{r=1}^{n}\sum_{\underset{\left(r\ne q\right)}{q=1}}^{n}\left(\prod_{\underset{\left(p\ne r,q\right)}{p=1}}^{n}\bar{R}_{\mu_{2p-1}\mu_{2p}}^{\nu_{2p-1}\nu_{2p}}\right)R_{\mu_{2r-1}\mu_{2r}}^{\nu_{2r-1}\nu_{2r}}R_{\mu_{2q-1}\mu_{2q}}^{\nu_{2q-1}\nu_{2q}}\\
=n\left(n-1\right)\delta_{\nu_{1}\dots\nu_{2n}}^{\mu_{1}\dots\mu_{2n}}R_{\mu_{1}\mu_{2}}^{\nu_{1}\nu_{2}}R_{\mu_{3}\mu_{4}}^{\nu_{3}\nu_{4}}\left(\prod_{p=3}^{n}\bar{R}_{\mu_{2p-1}\mu_{2p}}^{\nu_{2p-1}\nu_{2p}}\right),
\end{multline}
 after renaming the dummy indices and using the totally antisymmetric
nature of $\delta_{\nu_{1}\dots\nu_{2n}}^{\mu_{1}\dots\mu_{2n}}$.
Then, employing the (A)dS background Riemann tensor form (\ref{eq:AdS})
and using (\ref{eq:Recursion}), one gets 
\begin{multline}
\delta_{\nu_{1}\dots\nu_{2n}}^{\mu_{1}\dots\mu_{2n}}\sum_{r=1}^{n}\sum_{\underset{\left(r\ne q\right)}{q=1}}^{n}\left(\prod_{\underset{\left(p\ne r,q\right)}{p=1}}^{n}\bar{R}_{\mu_{2p-1}\mu_{2p}}^{\nu_{2p-1}\nu_{2p}}\right)R_{\mu_{2r-1}\mu_{2r}}^{\nu_{2r-1}\nu_{2r}}R_{\mu_{2q-1}\mu_{2q}}^{\nu_{2q-1}\nu_{2q}}\\
=n\left(n-1\right)\left[\frac{4\Lambda}{\left(D-1\right)\left(D-2\right)}\right]^{n-2}\frac{\left(D-4\right)!}{\left(D-2n\right)!}4\chi_{\text{GB}}.
\end{multline}
 With this result and (\ref{eq:First_order_contractions}), (\ref{eq:Zeroth_order_contractions}),
the second order term of the equivalent quadratic action becomes 
\begin{multline}
\left[\frac{\partial^{2}\mathcal{L}_{\text{Lovelock}}}{\partial R_{\rho\sigma}^{\mu\nu}\partial R_{\alpha\beta}^{\lambda\gamma}}\right]_{0}\left(R_{\rho\sigma}^{\mu\nu}-\bar{R}_{\rho\sigma}^{\mu\nu}\right)\left(R_{\alpha\beta}^{\lambda\gamma}-\bar{R}_{\alpha\beta}^{\lambda\gamma}\right)\\
=4\sum_{n=0}^{\left[\frac{D}{2}\right]}a_{n}n\left(n-1\right)\left[\frac{4\Lambda}{\left(D-1\right)\left(D-2\right)}\right]^{n-2}\frac{\left(D-4\right)!}{\left(D-2n\right)!}\chi_{\text{GB}}\\
-2\sum_{n=0}^{\left[\frac{D}{2}\right]}a_{n}n\left(n-1\right)\left[\frac{4\Lambda}{\left(D-1\right)\left(D-2\right)}\right]^{n-1}\frac{2\left(D-2\right)!}{\left(D-2n\right)!}\left(R-\frac{D\Lambda}{D-2}\right).
\end{multline}

\end{document}